\documentclass{jpp}
\usepackage{graphicx}

\usepackage[utf8]{inputenc}
\usepackage[T1]{fontenc}
\usepackage{amsmath}

\usepackage{graphicx}
\usepackage{dcolumn}
\usepackage{bm}
\usepackage{xcolor}
\usepackage[colorinlistoftodos]{todonotes}
\usepackage{soul}

\shortauthor{M.~Luo et al}

\title{Evolution of the autoresonant plasma wave excitation in two-dimensional particle-in-cell simulations}

\author{M.~Luo\aff{1}
  \corresp{\email{mufei@chalmers.se}}, C.~Riconda\aff{2}, A.~Grassi\aff{2},  N.~Wang\aff{3}, J.~S.~Wurtele\aff{4}, T.~F\"ul\"op\aff{1}, and I.~Pusztai\aff{1}}

\affiliation{\aff{1}Department of Physics, Chalmers University of Technology,  G\"{o}teborg, SE-41296, Sweden
\aff{2}LULI, Sorbonne Université, CNRS, École Polytechnique, CEA, 75252 Paris, France
\aff{3}College of Electrical Engineering, Zhejiang University, Hangzhou 310027, China
\aff{4}Department of Physics, University of California, Berkeley, California 94720, USA}

\begin{document}

\maketitle

\begin{abstract}

The generation of an autoresonantly phase-locked high amplitude plasma waves to the chirped beat frequency of two driving lasers is studied in two dimensions using particle-in-cell simulations.   
The two-dimensional plasma and laser parameters correspond to those that optimized the plasma wave amplitude in one-dimensional simulations. 
Near the start of autoresonant locking, the  two-dimensional simulations  appear similar to one-dimensional particle-in-cell results  [Luo et al., Phys.~Rev.~Res.~6, 013338 (2024)] with plasma wave amplitudes above the Rosenbluth-Liu limit. Later, just below wave-breaking, the two-dimensional simulation exhibits  a Weibel-like instability and  eventually laser beam filamentation. These limit the coherence of the plasma oscillation after the peak plasma wave field is obtained. In spite of the reduction of spatial coherence of the accelerating density structure, the acceleration of self-injected electrons in the case studied remains at $70\%$ to $80\%$ of that observed in one dimension. Other effects such as plasma wave bowing are discussed. 

\end{abstract}

\section{Introduction \label{introduction}}

Plasma beat-wave acceleration (PBWA), first proposed by Tajima and Dawson~\citep{Tajimaprl1979}, is based on driving relativistic plasma waves by the ponderomotive force of the beat wave of two -- typically picosecond long -- laser pulses. 
Recently there has been a renewed interest in the PBWA scheme, as an alternative to the prevailing laser wakefield acceleration (LWFA) scheme driven by femtosecond-pulses \citep{leemans2006gev,ke2021,oubrerie2022controlled,PhysRevLett.128.164801,zhu2023}, as it allows for efficient acceleration over a wider range of plasma and laser parameters. For instance, the PBWA can operate at relaxed requirements on laser diffraction \citep{PhysRevAccelBeams.26.061301}, allows electron acceleration at near-critical densities \citep{photonics9070476}, favors self-injection, and it can be combined with a plasma channel to control the phase velocity of the plasma wave \citep{plasma6010003}.  
Our previous kinetic study of this scheme~\citep{luo2024control} in one dimension (1D) demonstrates that autoresonance~\citep{Fajansajp2001, lindbergprl, lindbergpop, Chapmanprl, Yaakobi2008, Luopop} can increase the plasma wave amplitude beyond the Rosenbluth-Liu (RL) limit~\citep{Rosenbluthprl1972}, up to the wave-breaking limit, and it provides guidance to choose the laser and plasma parameters optimally. However, the desirable properties of autoresonant PBWA that survive the test of multi-dimensionality remain to be demonstrated.

In this paper, we employ the fully kinetic, two-dimensional (2D) particle-in-cell (PIC) code {\sc Smilei} \citep{DEROUILLAT2018351} to investigate multi-dimensional effects on autoresonantly-driven large amplitude plasma waves.  This extends previous work~\citep{luo2024control}, where we examined, again using  {\sc Smilei}, autoresonantly driven plasma wave excitation in one dimension under a wide range of plasma and laser parameters. In this paper, we restrict to a 2D  study of parameters that are representative in the 1D case of large amplitude plasma waves and electron self-trapping in the nonlinear regime.  We are focused on the plasma dynamics, including both nonlinear and 2D effects, in the presence of autoresonant excitation of a plasma wave. The autoresonant wakefield excitation process has not been previously studied in 2D with a kinetic code.

In some recent work~\citep{oxford1,oxford3,oxford2}, a two-pulse scheme was studied, with a short initial relatively high-intensity drive laser, followed by a short gap, and then a trailing relatively low-intensity laser.  Their scheme reaches the GeV-scale acceleration of an injected bunch over a distance of $\sim100\,\rm mm$. In our simulations, both pulses are long and low-power. We do not consider here the performance of the 2D autoresonant beat wave with an injected bunch. At the high fields studied in our paper, beyond the RL limit and up to wave-breaking, we see self-injection of electrons which are subsequently accelerated to $\sim 200\,\rm MeV$ over $\sim 3.5\,\rm mm$.  In comparison to~\citet{oxford1,oxford3,oxford2}, the peak longitudinal electric field is higher here, because we have a higher density and we exceed the RL limit, up to the point of wave breaking.

We find that in 2D, the acceleration of self-injected electrons remains comparable to that observed in 1D, particularly in the early stages of the process. Further back in the process, after the peak plasma wave amplitude is reached, a variety of  2D effects emerge \citep{forslund1985prl} that reduce the wave amplitude and transverse coherence. These include parametric coupling~\citep{Andreas,MfluoSRS},  Weibel-like instabilities~\citep{weibelInstability}, and laser beam filamentation~\citep{RevModPhys.81.1229}. These 2D effects result in a reduced acceleration efficiency for self-injection of $70$-$80\%$ of that in the 1D case.

The rest of the paper is structured as follows. In section~\ref{kinetic_simulation}, we present 2D kinetic simulations of autoresonant PBWA, with the simulation setup explained in section~\ref{setup}, the wave excitation is examined in section~\ref{overview} and accelerated electrons are discussed in section~\ref{energetic}. Weak nonlinearities are examined in section~\ref{EarlyStage}, a Weibel-like instability is identified in section~\ref{transitionStage}, and the strongly nonlinear regime and laser filamentation are studied in section~\ref{FinalStage}. The results are summarized and discussed in section~\ref{conclusion}.

 
\section{Autoresonant PBWA in two dimensions\label{kinetic_simulation}}

We consider the most important characteristics of plasma beat-wave acceleration in two dimensions. We shall find that the early stages of the wave excitation and acceleration processes do not deviate significantly from corresponding one-dimensional results, while at later times the acceleration in 2D starts to fall short of that in 1D. The detailed analysis of these deviations will be addressed later in section~\ref{phases}.

\subsection{Simulation setup\label{setup}}

In our 2D investigation of the autoresonant PBWA, the two co-propagating laser beams are chosen to have parallel linear polarization and identical intensities. They have the shape of a 6th-order super-Gaussian temporal profile ~$\propto\exp[-(t/T_{\rm pulse})^6]$, and ~$\propto\exp[-(y/w_0)^6]$ in the transverse ($y$) direction, where the width of the laser beam $w_0$ is set to be $2.4\pi k_{\rm pe}^{-1}$. Here, $k_{\rm pe}=\omega_{\rm pe}/c$ is the plasma wavenumber, $\omega_{\rm pe}=\sqrt{n_{e} e^2/\epsilon_0 m_e}$ is the electron plasma frequency,  $n_{e}$ is the background electron density, $-e$ and $m_e$ are the electron charge and mass, and $c$ is the speed of light in vacuum.

\begin{table}
  \begin{center}
\def~{\hphantom{0}}
  \begin{tabular}{ p{5cm} | p{5cm} }
\textbf{Parameter}                &                  \textbf{Value} \\[3pt]          
Laser wavelength                  &           800 nm     \\  [1.ex] 
Laser intensity                   &        $8.5\times10^{16}\;\rm W / cm^{2}$\\   [1.ex] 
Laser spot size                   &          $48 \rm \;\mu m$       \\  [1.ex] 
Laser duration                    &          $4.3 \;\rm  ps$       \\  [1.ex] 
Laser bandwidth ($\omega_1$)      &         $0.56\%$     \\  [1.ex] 
Plasma density                    &      $7\times10^{17}\;\rm cm^{-3}$ \\  [1.ex] 
Gradient length                   &          $16 \;\rm \mu m$ \\  [1.ex] 
Rosenbluth-Liu Limit              &        $47 \;\rm GV/m$    \\  [1.ex] 
Wave-breaking                     &        $80\; \rm GV/m$    \\                     
  \end{tabular}  
  \caption{Simulation parameters}
  \label{table_parameters}  
  \end{center}
\end{table}

The laser and plasma parameters are chosen based on our previous 1D study, in order to reach the wave-breaking limit. Namely, the amplitudes of the two lasers (indicated by subscripts $1$ and $2$) are $a_1=a_2= 0.2$ in terms of the normalized vector potential $a = e A/m_ec$. The ratio of homogeneous plasma density $n_{e}$ and the critical density $n_{cr}$ is $0.0004$, $n_{cr}=\omega_{1}^2m_e\epsilon_0/e^2$ is the critical density corresponding to a laser frequency $\omega_1$, with $\omega_1$ the central frequency of the beam one (1). Before entering the homogeneous plasma, a linear density ramp with the gradient length $L_{\rm ramp}k_{\rm pe}=40\pi$ is applied. A chirp rate $\alpha = -0.0014$ is applied to the first laser beam, giving the frequency difference of the two laser beams 
$\Delta\omega=\omega_{\rm pe}[1+\alpha(t-t_0)\omega_{\rm pe}]$. At time $t_0=22.5\pi/\omega_{\rm pe}$, the frequency difference between the two lasers matches the nominal plasma frequency, $\Delta\omega=\omega_1-\omega_2=\omega_{\rm pe}$. Here the corresponding laser frequencies are $\omega_1/\omega_{\rm pe}=50$ and $\omega_2/\omega_{\rm pe}=49$. The laser duration $T_{\rm pulse}\omega_{\rm pe}=64\pi$ is chosen to promote the autoresonant growth of the plasma wave until the wave-breaking limit. Further details on the choice of these parameters are given in Appendix~\ref{control_duration}. The ions are set to be immobile, as they are not essential for the dynamics on the time scale of the laser pulse passing by~\citep{Mora1988prl}. The longitudinal and transverse spatial resolutions are $dx=0.008k_{\rm pe}^{-1}$ and $dy=2dx$, respectively. Table~\ref{table_parameters} shows the experimentally relevant parameters for a standard chirped pulse amplified $800\,\rm nm$ Ti:sapphire laser.

\begin{figure}
    \centering
	\includegraphics[width=0.75\linewidth]{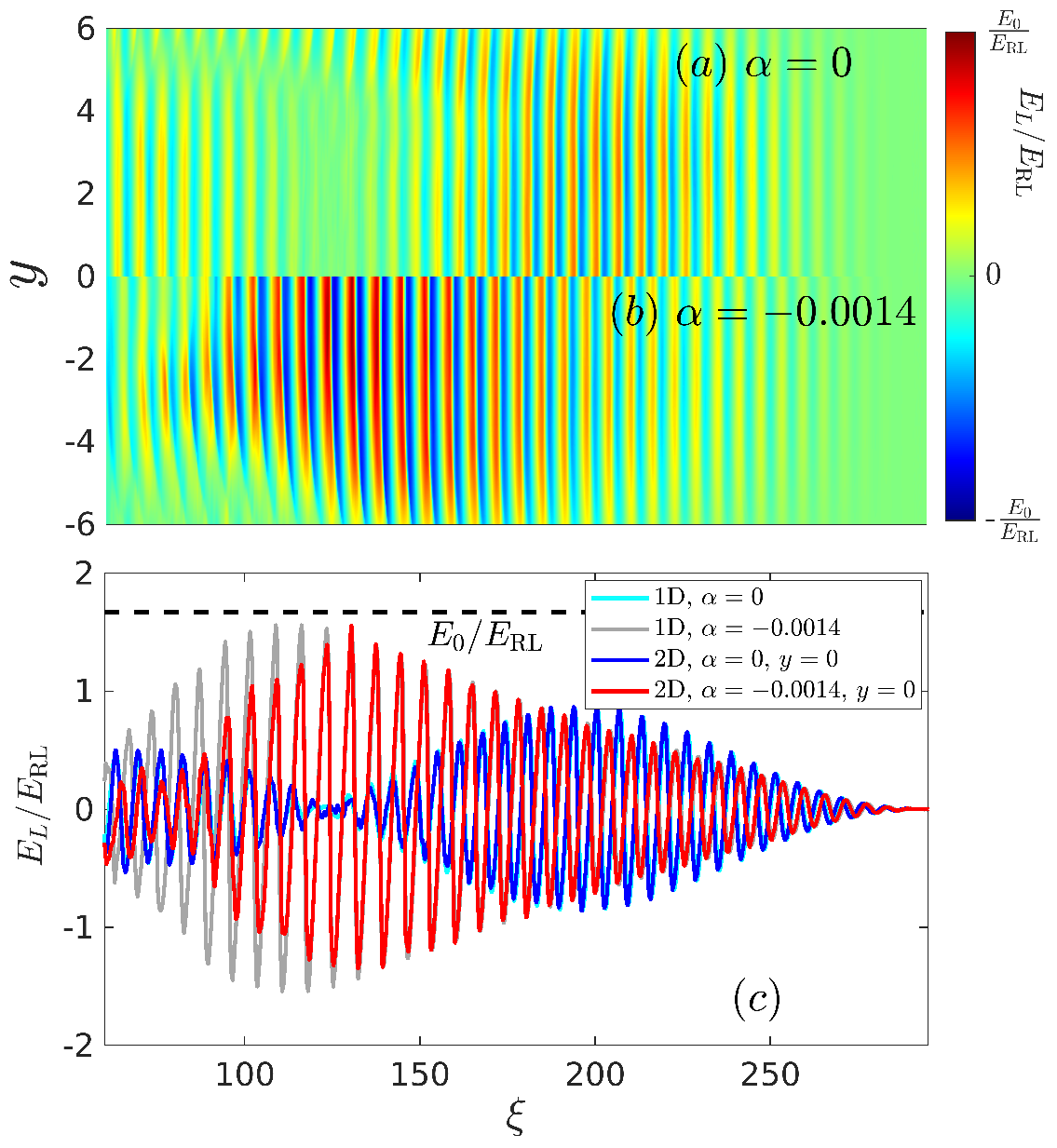}
    \caption{The longitudinal electric field of the plasma wave normalized to the RL field, $E_L / E_{\rm RL}$, in the $2 \mathrm{D}$ simulations, evaluated at $t \omega_{\mathrm{pe}} \approx 250$. The duration of the laser pulses is $T_{\text {pulse }} \omega_{\text {pe }} \approx 64 \pi$. (a) No chirp. (b) With chirp rate $\alpha=-0.0014$. (c) Electric field at the symmetry axis ($y=0$). The blue (red) line corresponds to simulations without (with) chirp in 2D. Corresponding 1D simulation data are shown by the light-blue (gray) lines without (with) chirp. The black dashed line indicates the wave-breaking field.}
\label{2D_electric_field} 
\end{figure}

\begin{figure}
    \centering
	\includegraphics[width=1\linewidth]{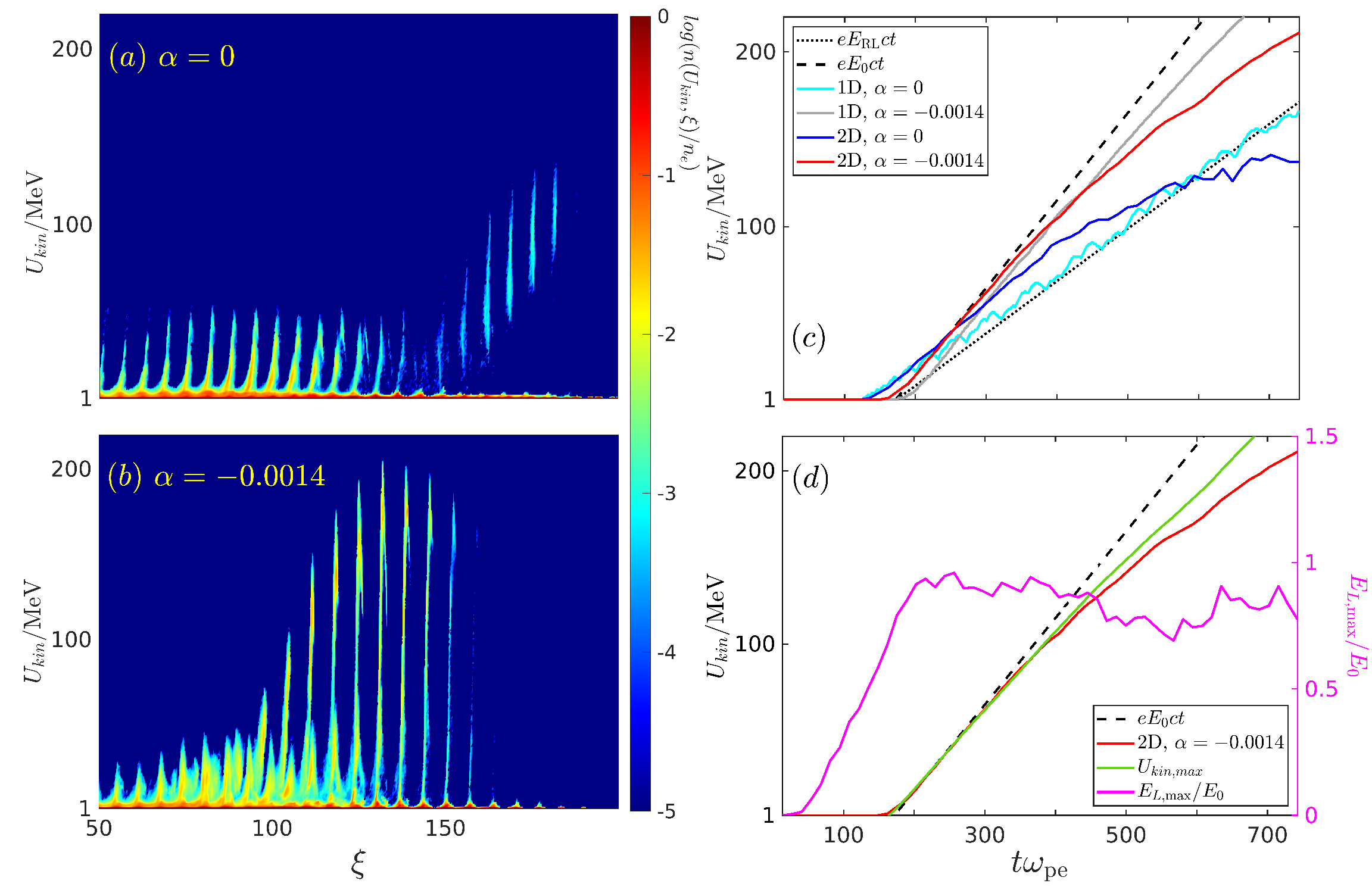}
    \caption{The kinetic energy spectrum of electrons as a function of the co-moving coordinate $\xi$, integrated over $y$, at $t \omega_{\mathrm{pe}} \approx 750$ are plotted in (a) without chirp and (b) with chirp rate $\alpha=-0.0014$. (c) Time traces of the highest electron kinetic energy for a variety of cases. The red ($\alpha=-0.0014$) line and blue line ($\alpha=0$) are in $2 \mathrm{D}$, the gray ($\alpha=-0.0014$) line and light blue line ($\alpha=0$) are in 1D. The black dashed and dotted lines indicate the kinetic energies of an electron that was accelerated by a constant electric fields at the wave-breaking ($E_0$) and the RL limit ($E_{\rm RL}$). (d) The temporal evolution of the maximum electric field $E_{L, \max }$ is shown in magenta (right axis). If an electron was accelerated with this time-varying field, it would reach energies as shown by the green curve (left axis). For reference the dashed and the red lines of panel (c) are repeated.}  
\label{2D_energy_spectrum} 
\end{figure}

\subsection{Autoresonant plasma wave excitation in 2D \label{overview}}

First we consider the growth phase of the autoresonant beat-wave excitation, comparing 1D and 2D results. Figure~\ref{2D_electric_field} shows the longitudinal electrostatic field $E_L$ in the 2D simulation at time $t \omega_{\rm pe}\approx 250$. The electric field is normalized to the Rosenbluth-Liu limit $E_{\rm RL}=(16a_1a_2/3)^{1/3}E_0$, where $E_0=m_ec\omega_{pe}/e$ is the cold, nonrelativistic wave-breaking field. The longitudinal coordinate is the co-moving position $\xi=\omega_{\rm pe}(x/c-t)$.  Figures~\ref{2D_electric_field} (a) and (b) correspond to simulations without chirp and with a chirp rate $\alpha=-0.0014$, respectively, that was deemed optimal for field enhancement with these parameters. {The field enhancement with respect to the zero chirp case is evident in figure~\ref{2D_electric_field} (a) and (b). }
The maximum value of the electric field in the chirp case is close to the cold, non-relativistic wave-breaking threshold $E_0$ that takes the value $E_0/E_{\rm RL}\approx 1.67$. 

Notice that as clearly visible in figure~\ref{2D_electric_field}(a),  for  the case $\alpha=0$ the plasma wave amplitude  is larger off-axis than closer to the symmetry axis, in the region $\xi\approx 100$-$150$. This is caused by the finite size of the laser field in the transverse direction and the  related intensity variation. As a consequence the growth and saturation time of the plasma wave excitation varies transversally. Specifically, the saturation time in the standard PBWA scheme is approximately $t_{\rm sa}\omega_{\rm pe}=3.68(\omega_{\rm pe}\tau_{r})^{4/7}(a_1a_2)^{-2/7}$~\citep{forslund1985prl}, where $\tau_{r}$ is the rise-time of the laser beams' temporal profile, introduced in section~\ref{setup}, leading to slower saturation of the plasma wave off-axis. Eventually, the RL-limit is also reached at the off-axis regions, as visible in the plot.

Figure~\ref{2D_electric_field}(c) shows the electric field variation along the symmetry axis, i.e., $E_L(\xi, y=0)$. The blue (red) line corresponds to the case without (with) chirp in 2D. For reference, we also show the 1D simulation results, shown by the light-blue (gray) line without (with) chirp. The no-chirp results in 1D and 2D overlap, and show that the RL value is indeed the limit of the achieved electric field; the maximum of the blue curves is at $E_L/E_{RL}=1$. When the optimal chirp value of $\alpha=-0.0014$ is employed, the autoresonant growth of the plasma wave is achieved and $E_L/E_{RL}=1$ is exceeded, but the overlap of the 2D and 1D results (red and gray, respectively) extends only for $\xi \gtrsim 125$. Below this value the 2D results start to deviate from the 1D ones. The reasons for this are discussed in detail in section~\ref{phases}.

\subsection{Energetic electron dynamics \label{energetic}}

After concluding that autoresonant plasma wave excitation is possible in 2D, we will now turn to the electron acceleration aspect of the process. 
Figure~\ref{2D_energy_spectrum}(a) and (b) show the electron phase space in terms of the co-moving coordinate $\xi$ and the kinetic energy reached by the electrons $U_{kin}$ (the data is integrated over the transverse coordinate). Self-injected electrons can be effectively accelerated by the electric field structure of the plasma wave, and as expected, electrons can reach higher energies -- in this case $200\,\rm eV$ at $t\omega_{\rm pe}\approx 750$ -- when a finite chirp of $\alpha=-0.0014$ is employed, allowing for autoresonant field growth; compare figure~\ref{2D_energy_spectrum}(b) to (a), where $\alpha=0$. {Notice that in the chirped case electrons are mainly accelerated in the region between $100< \xi < 150$, while in the region  between $50 < \xi < 100$ we observe energetic particles but at lower energies. This was also observed in the analogous 1D kinetic simulation~\citep{luo2024control}, and can be understood in the following way. Because of autoresonance the plasma wake field grows with decreasing $\xi$. In the fluid model for the parameters chosen here (see Appendix~\ref{control_duration}) it can become larger than $E_0$, but in kinetic simulation, once it reaches $E_0$, it starts to trap very efficiently particles that in turn stop the autoresonance. As a result the field is highest between $100< \xi < 150$ with value slightly below $E_0$. This is clearly visible in figure~\ref{2D_electric_field}(c) for both the 2D and 1D chirped simulations (gray and red curves), the decrease of the field for $\xi < 100$ being more pronounced in the 2D case.}

{We now focus on the region $\xi> 100$}. To assess the efficiency of the acceleration process, we track the kinetic energy of electrons reaching the highest energies (note that the acceleration starts at slightly different times in the various simulations). In figure~\ref{2D_energy_spectrum}(c) we compare the temporal evolution to estimate the acceleration efficiency corresponding to the acceleration of relativistic electrons by a constant electric fields at the RL limit, $E_{\rm RL}$ (dotted line), and at the wave breaking value, $E_0$ (dashed line). As expected, we find that in the no-chirp cases, the kinetic energy mostly increases with a steepness dictated by the RL field limit. The 1D simulation (light blue) stably accelerates at this rate up to the end of the considered time range, while we observe a reducing slope in 2D (dark blue). In the simulations with chirp, the electron energy initially increases closely following the wave-breaking field estimate. At later times, the slope changes slightly in 1D and more significantly in the 2D simulation. The 1D and 2D autoresonant results (gray and red line, respectively) quite closely follow each other up to $t\omega_{\rm pe}\approx 450$, indicating a similar performance in the early stage of the acceleration.

Two reasons can be responsible for the difference between the autoresonant simulations and the estimate indicated by the dashed line. First, the maximum electric field might not be sustained at the constant level of $E_0$, and second, the particles are not necessarily positioned to be optimally accelerated by the maximum electric field. We find that the first effect is the one mainly  at play. Indeed the kinetic energy of the most energetic particle from the simulations $ U_{kin, \rm sim}$ as function of time should follow $U_{kin, \rm max}$ defined by 

\begin{equation}
    U_{kin, \rm max}(t)=\int_{t_0}^t eE_{L,\rm max}(t')cdt',
    \label{kinetic_energy_sim}
\end{equation}
where $E_{L, \rm max}$ is the maximum instantaneous electric field. In figure~\ref{2D_energy_spectrum}(d), the time evolution of $E_{L, \rm max}$ is shown (magenta, right axis) along with the corresponding $U_{kin, \rm max}$ (green) and $U_{kin, \rm sim}$ (red). While $E_{L, \rm max}$ remains in the vicinity of $E_0$, it is somewhat lower most of the time. Thus, the expected particle energy (green curve) departs downward from the theoretical upper bound (dashed). The actual electron energy (red curve), closely follows $U_{kin, \rm max}$ until $t\omega_{\rm pe}\approx 400$, and then starts to deviate. This marks the beginning of a 2D effect of the system that is strong enough to degrade the efficiency of acceleration. The responsible physical processes are to be detailed in section~\ref{phases}. 

\begin{figure}
    \centering
	\includegraphics[width=0.75\linewidth]{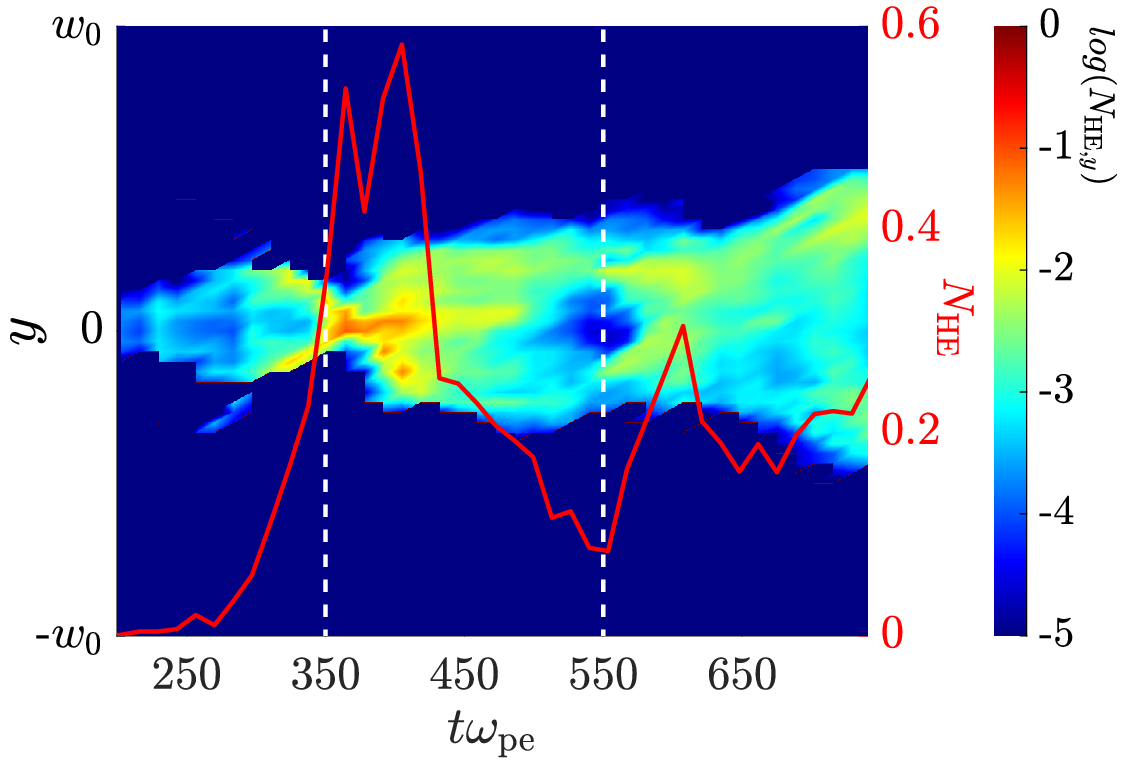}
    \caption{Color plot: The high-energy electron distribution  ($N_{\rm HE,y}(y,t)$) is plotted. Red curve (right axis):  $N_{\rm HE}(t)$ (integrating over the $y$-axis). The two vertical dashed lines mark the times considered in figure~\ref{high_energy}, when the energetic electrons are localized on axis and off-axis, respectively.}  
\label{centering} 
\end{figure}

Next, we consider the transverse distribution of high-energy particles.
Figure~\ref{centering} shows the transverse distribution of the most energetic electrons as a function of time. At each time electrons in the uppermost $10\%$ of the currently spanned energy range are identified and their transverse distribution is color plotted. The transverse integral of this color plot is shown as red curve (right axis). More specifically, defining the "high-energy" electron density by $n_{\rm HE}(\xi,y,t)=\int_{0.9 U_{kin,{\rm max}}(t)}^{U_{kin,{\rm max}}(t)}f_e(U_{kin},\xi,y,t)dU_{kin}$, with the electron distribution function $f_e$, the color plot shows $N_{\rm HE,y}(y,t)=\int d\xi\, n_{\rm HE}(\xi,y,t)/n_e$, with the nominal electron density $n_e$, while the red line is $N_{\rm HE}(t)=\int dy \,N_{\rm HE,y}(y,t)$; note that here the $y$ coordinate is normalized to $k_{\rm pe}^{-1}$ (similarly to $\xi$). When interpreting this image, we must keep in mind that the considered energy range corresponds to continuously increasing energies, since $U_{kin,{\rm max}}=U_{kin,{\rm max}}(t)$. The highest number of energetic electrons is reached around $t\omega_{\rm pe}\simeq 400$, then it rapidly drops {and stays around $20\%$. After this time the electron energy keeps increasing but for a smaller population than at earlier times}. { Up to  $t\omega_{\rm pe}=350$  the fast electrons are mostly  localized at the center, while we observe  a rapid widening of the transverse distribution after that}. Even later, around  $t\omega_{\rm pe}=550$, the most energetic electrons are found off-axis.

\begin{figure}
    \centering
	\includegraphics[width=0.75\linewidth]{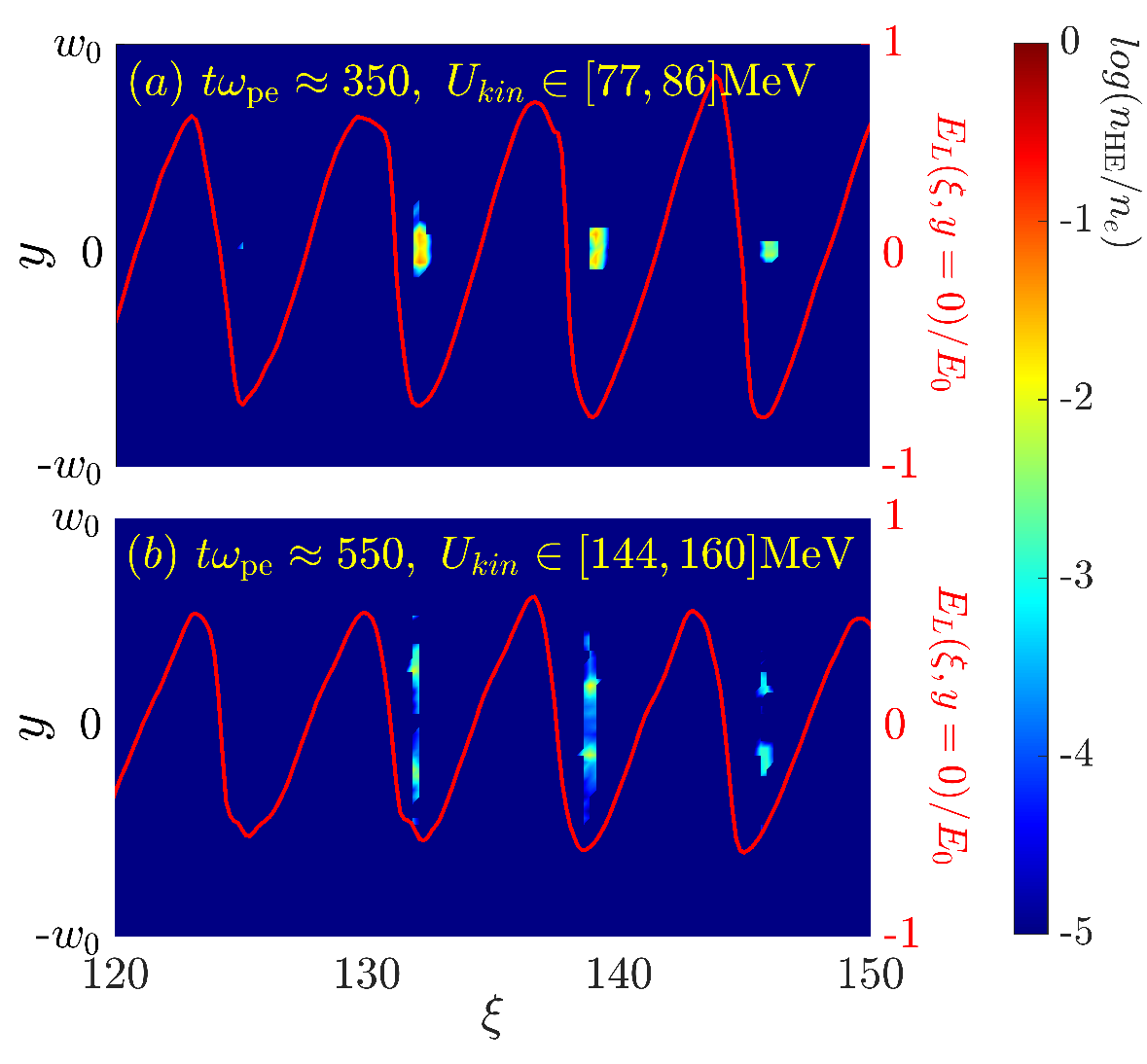}
    \caption{Spatial distribution of $n_{HE}(\xi,y)$ at (a) $t\omega_{\rm pe}\approx 350$ and (b) $t\omega_{\rm pe}\approx 550$. These times are marked with dashed lines in figure~\ref{centering}. The longitudinal electric field at the axis, $E_L(\xi, y=0)/E_0$, is indicated by the red curves (right axes).}  
\label{high_energy} 
\end{figure}

We now consider two snapshots, at the times indicated by the white dashed vertical lines in figure~\ref{centering}, and we show in figure~\ref{high_energy} the logarithm of the relative number of energetic electrons as function of $\xi$ and $y$. At $t\omega_{\rm pe}=350$, shown in figure~\ref{high_energy}(a), the maximum kinetic energy is $U_{kin,{\rm max}}=86\,\rm MeV$, and as expected 
the energetic electron bunches are close to the symmetry axis. At the later time,  $t\omega_{\rm pe}=550$, shown in figure~\ref{high_energy}(b), the distribution peaks off-axis, and the instantaneous maximum energy is $U_{kin,{\rm max}}=160\,\rm MeV$. The bunches are localized at the minimum of the electric field where the acceleration is the strongest, indicated by the red curves in the same figures. In the following section, we will delve into the reasons for the transverse dynamics, which shall also shed light on why the acceleration efficiency is reduced when moving to one to two dimensions.

\section{Phases of intensifying nonlinear effects \label{phases}}

The regime explored in this work corresponds to the beat-wave autoresonance excitation of very large amplitude plasma waves  and significant particle trapping, resulting in a number of 2D nonlinear processes that affect the evolution of the plasma wave. It is interesting however, that even in these extreme conditions the nonlinearities do not destroy the laser-plasma coupling and the particle acceleration, but only partially degrade the process. Here we review these effects, putting the focus on strong nonlinear effects appearing late in time, which impact the temporal coherence of the plasma wave. The understanding of these processes can provide guidance for future experiments, where they may be reduced or eliminated through appropriate schemes. 

\subsection{The weakly nonlinear stage \label{EarlyStage}}

In the early phase of the evolution of the plasma wave -- apart from a downward shift of its wavenumber as the plasma wave amplitude increases -- the wave-number spectrum is affected by {forward Raman scattering at a slight angle with respect to the laser propagation. This leads to the appearance of a double signal in the electromagnetic spectrum, mainly around $k_L/k_{\rm pe} = 50$~\citep{forslund1985prl}, and a broadening of both the electromagnetic and the electron plasma wave spectrum in the transverse direction. Moreover, Stokes and anti-Stokes scattering results in harmonics  of the electromagnetic wave. In Appendix~\ref{scatterings} we plot the Fourier spectrum of the waves, and all these components are clearly visible. At this stage, the only new effect compared to the 1D case, discussed in \citet{luo2024control}, is the transverse broadening and the double signal around  $k_s/k_{\rm pe} = 50$, while the Stokes and anti-Stokes components and the longitudinal widening were already present in the 1D simulations. Overall the autoresonant  excitation and the particle acceleration appear robust with respect to the presence of the transverse Raman scattering.} 

Another 2D effect that can be identified, is due to the transverse variation of the laser intensity, and the corresponding nonlinear correction to the wavelength. This leads to the formation of slightly curved plasma wave fronts, while the plasma wave structure remains coherent, as was also observed in some stimulated Raman scattering simulations~\citep{PEM, Lin1,Lin2}. More details on this \emph{wavefront bowing} are given in Appendix~\ref{bowing}. Similarly to the mentioned parametric wave coupling processes, the wavefront bowing effect has no appreciable negative effect on the efficiency of the autoresonant PBWA.

\subsection{Transverse magnetic field growth due to a Weibel-like instability \label{transitionStage}}

We now focus our attention on processes that become important around the time the acceleration efficiency starts to degrade compared to the 1D simulations. The first of these effects is the growth of a Weibel-like instability \citep{weibelInstability,Fried1959}, which provides a seed for subsequent plasma wave filamentation \citep{PEM}.  

\begin{figure}
    \centering
	\includegraphics[width=0.75\linewidth]{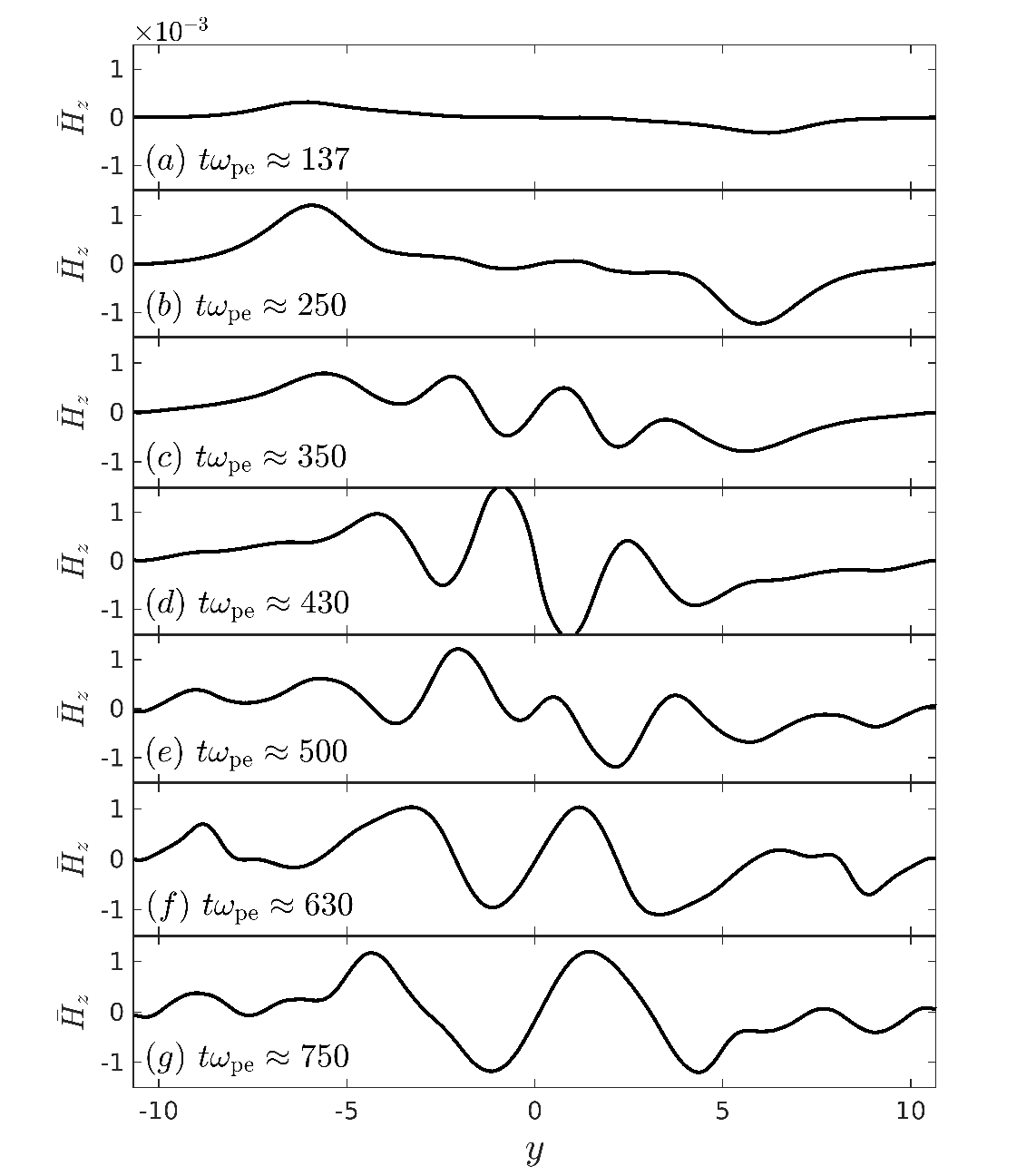}
    \caption{Longitudinally averaged transverse profiles of the $z$-component of the magnetic field, taken at different times, normalized as $\bar{H}_z=e\bar{B}_z/m_e\omega_1$. The averaging is performed over the same $\xi$-range as covered by the dashed rectangle in figure~\ref{weibel}(a).}  
\label{Bz_times} 
\end{figure}

\begin{figure}
    \centering
	\includegraphics[width=0.75\linewidth]{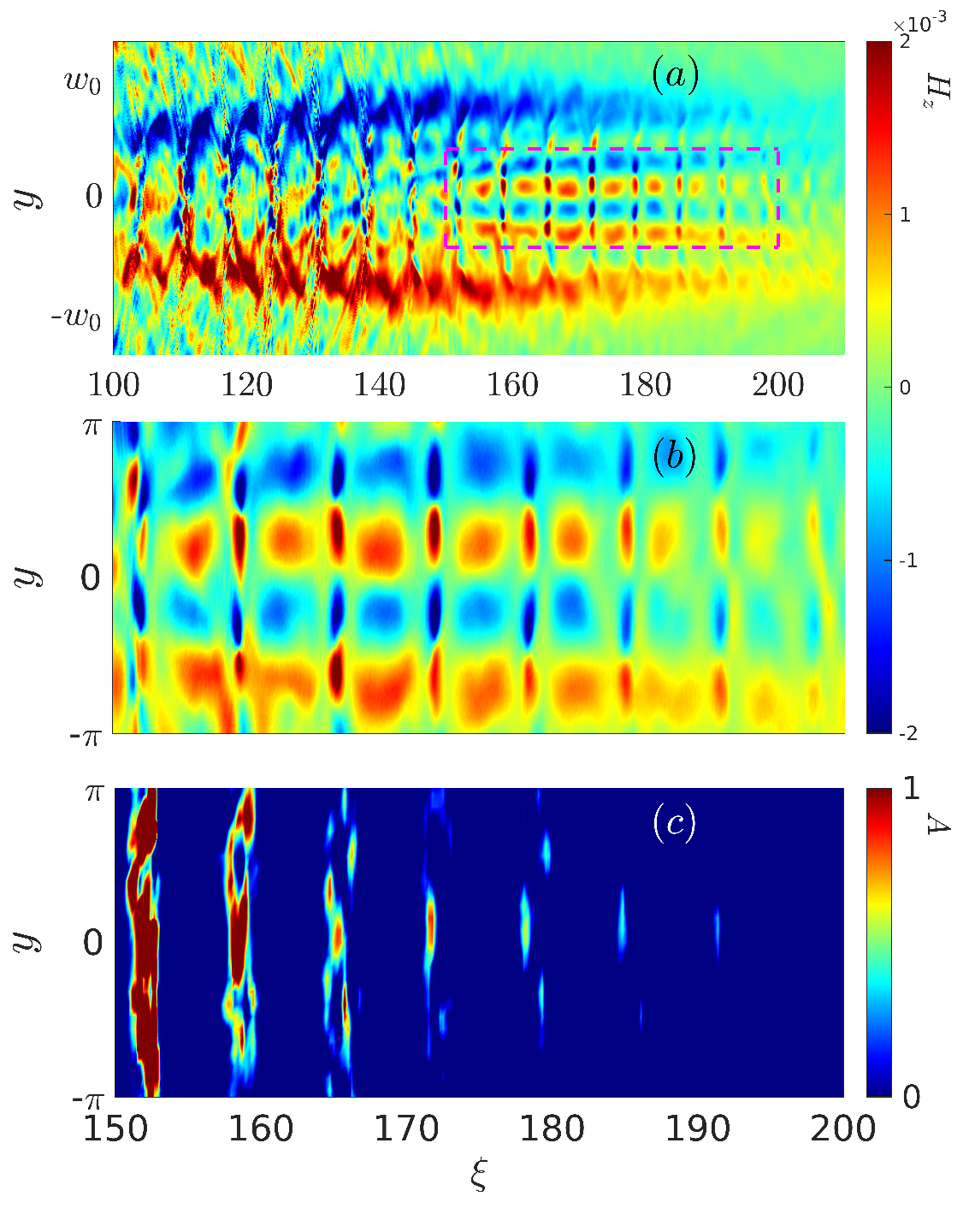}
    \caption{(a) Spatial variation of the magnetic field $H_z=eB_z/(m_e\omega_1)$ at time $t \omega_{\rm pe}\approx 350$. The filamentation of the magnetic field is clearly visible inside the magenta dashed box, where the plasma wave is driven to significant amplitudes while wave-breaking is not reached. (b) Zoom-in of the magnetic field, taken from the region highlighted in panel (a). (c) Temperature anisotropy $A$ in the same region as panel (b). Note that for the visibility, the range of $A$  is capped at $1$, while $A$ can reach values of $\approx 3$ but only in very restricted regions. }  
\label{weibel} 
\end{figure}

In figure~\ref{Bz_times}, the longitudinally averaged transverse magnetic field $\bar{B}_z$ is shown as a function of the transverse coordinate $y$, normalized as $\bar{H}_z=e\bar{B}_z/m_e\omega_1$. The times shown illustrate the development of transverse magnetic field filaments (time increases from panel (a) to (g)). The average in the longitudinal direction is performed over the $\xi$-range indicated by the dashed rectangle in figure~\ref{weibel}(a). In this region, the plasma wave has a significant amplitude, but wave-breaking has not yet been reached, and only a few particles are trapped. Indeed, the $\xi - p_x$ phase-space (not shown here) has a very regular structure with longitudinal velocity up to $\simeq c$. Stronger particle acceleration is instead observed in the range $100< \xi < 150$, leading to a more disordered magnetic field structure. 
In the early stages shown in figure~\ref{Bz_times}(a-b), the magnetic field starts growing at the edges of the laser beams ($y\sim\pm 2.4\pi$) because of the large-scale current generated by the density and velocity perturbation in the plasma wave (with a homogeneous component plus one varying in the longitudinal direction with wavenumber $\sim 2 k_{\rm pe}$), as discussed by ~\citet{gorbunov1,gorbunov2,sheng7}. Some filaments start to appear in the central region and become more pronounced by the time shown in figure~\ref{Bz_times}(c).
As expected for a Weibel-like instability, the filaments grow with a typical wavenumber of the order of $k_{\rm pe}$. Both wavelength and field amplitude increases with time (figure~\ref{Bz_times}(d)-(g)) up to saturation, happening at $\omega_{\rm pe}t \simeq 430$. 

To understand the origin of these magnetic field filaments, figure~\ref{weibel}(a) shows the magnetic field $H_z=eB_z/(m_e\omega_1)$ as a function of $\xi$ and $y$ at time $t\omega_{\rm pe}\approx 350$ (corresponding to the longitudinally averaged result of figure~\ref{Bz_times}(c)). 
The transverse limits of the dashed magenta rectangle are located at $y=\pm\pi$ ($y$ is normalized by the plasma wavenumber $k_{\rm pe}$), and are chosen such that the rectangle covers the flat-top region of the transverse profile of the laser beams.
Inside this marked area, magnetic field filaments are clearly visible, as highlighted by figure~\ref{weibel}(b) which shows a zoom-in on the same data.

To identify the drive of the instability and estimate its growth rate, we define the longitudinal and transverse thermal speeds $v_{tx}$ and $v_{ty}$, respectively, as follows
$    v_{tj}^2(\xi,y)=n^{-1}\int (v_{j}-V_j)^2 f(\xi,y,\mathbf{v})d^3v$,
where $n=\int f(\xi,y,\mathbf{v})d^3v$, $V_j=n^{-1}\int v_j f(\xi,y,\mathbf{v})d^3v$, and $j=\{x,y\}$. These definitions allow us to introduce a temperature anisotropy parameter $A(\xi,y)=(v_{tx}^2/v_{ty}^2)-1$, adapting previous definitions~\citep{Ruyer2015,SilvaT, Silva2021} to our conditions. Its values are plotted in figure~\ref{weibel}(c) at $t\omega_{\rm pe}\approx 350$.
Since most of the acceleration happens longitudinally, the longitudinal spread of the distribution function and hence the thermal speed is higher than the transverse one, yielding a 
temperature anisotropy with values $A>0$. Significant values of $A$ up to $\approx 3$ are reached in patches, as shown in figure~\ref{weibel}(c). This anisotropy makes the plasma unstable to a Weibel-like instability, leading to the growth of the magnetic filaments.

\begin{figure}
    \centering
	\includegraphics[width=0.75\linewidth]{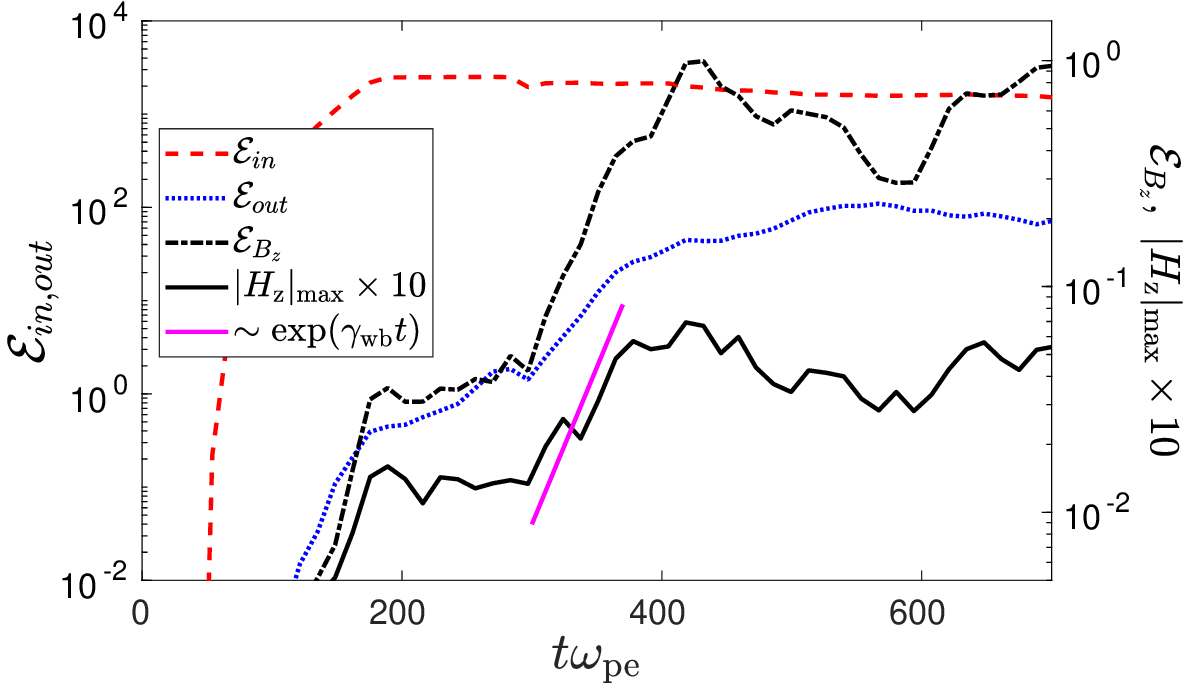}
    \caption{Maximum of the z-component of the magnetic field (solid black curve, $10 |H_z|_{\rm max}$), and energy contained in the z-component of the magnetic field (black dash-dotted, $\mathcal{E}_{B_z}$); both within the region indicated by the dashed magenta rectangle in figure~\ref{weibel}(a). The solid magenta line indicates an exponential growth with the estimated growth rate of the Weibel-like instability. The red dashed line is the quasi-longitudinal electric field energy $\mathcal{E}_{\rm in}$, and the blue dotted line is the electric field energy scattered outside the quasi-longitudinal wavenumber region $\mathcal{E}_{\rm out}$.}  
\label{Bz_energy} 
\end{figure}

The time evolution of the maximum value of $|H_z|$ and the magnetic energy content inside the marked rectangular region of figure~\ref{weibel}(a), $\mathcal{E}_{B_z}=\int\int{|eB_z/(m_e\omega_1)|^2}d\xi dy$, are shown in figure~\ref{Bz_energy}, with solid and dash-dotted lines, respectively. We choose to focus only on the magnetic field inside the marked region, where the Weibel-like instability is the dominant effect that generates the magnetic filaments. 
We hence neglect the field growing at the intensity gradients of the laser beams and the more complex structures in the rear region ($100<\xi < 140$), where efficient acceleration occurs near the axis, modifying the plasma wave structures and the phase space in a non-trivial way. This allows us to identify the typical growth of the field generated by the Weibel-like instability, as discussed below.

The general trends of $\mathcal{E}_{B_z}$ and $|H_z|_{\rm max}$ are similar, and reflect the observations from figure~\ref{Bz_times}. A first early rapid increase and saturation is observed at $t \omega_{\rm pe}\approx 190$. This initial phase of magnetic field generation is caused by the development of the self-generated quasi-static magnetic field as stated in Refs.~\citet{gorbunov1,gorbunov2,sheng7}. This is followed by another significant growth until $t\omega_{\rm pe}\approx 400$, after which the magnetic field saturates around a higher level, with some oscillations. This second growth phase corresponds to the Weibel-like instability. The maximum growth rate of this Weibel-like instability, $\gamma_{\rm wb}$, in terms of the anisotropy parameter above defined, is approximately given by~\citep{okada_yabe_niu_1978,KazuhitoSatou_1997, Sugie_2006, Okada2007pop, Zhouyz2022}
\begin{equation}
    \gamma_{\rm wb}/\omega_{\rm pe}=\sqrt{\frac{8}{27\pi}}\frac{v_{ty}A^{3/2}}{A+1}. 
    \label{wbGrowth}
\end{equation}
Considering values extracted from the simulation for the transverse thermal speed $v_{ty}\approx 0.1c$ and temperature anisotropy $A \approx 2.5$ -- where the latter is close to the upper end of the observed range of $A$ values -- yields a growth rate $\gamma_{\rm wb}/\omega_{\rm pe}\approx0.035$. Such an exponential growth, $\propto \exp(\gamma_{\rm wb}t)$, is indicated by the magenta line in figure~\ref{Bz_energy}, showing a satisfactory agreement with the growth of $|H_z|_{\rm max}$ (solid black curve).

In the weakly non-linear stage, discussed in detail in Appendix~\ref{scatterings}, we already observed that forward/side Raman scattering and wave bending  will tend to deform the plasma wave wavefront and introduce a transverse component to its wavenumber. The Weibel-like instability also affects the transverse wavenumber of the plasma wave, $k_{Ly}$. In order to quantify the effect of the Weibel-like instability on the transverse wavenumber of the plasma wave $k_{Ly}$, 
 we can take advantage of the fact that it will induce shorter transverse lengths (larger $k_{Ly}$) than the one already identified in the weakly non-linear stage. The longitudinal wavenumber $k_{Lx}$ instead is given by the matching condition,  roughly $k_{Lx}\sim k_{\rm pe}$, so we may consider a $k_{Lx}$ range that provides a comfortable margin around this\footnote{The harmonics of the plasma wave are not considered here, since they are negligible in our simulations.}, such as $k_{Lx}\in [0.5,1.5]k_{\rm pe}$.

 The characteristic transverse broadening of the wavenumber spectrum due to the Raman process can be estimated as  $\delta k_{Ly}=\sqrt{k_{\rm pe}^2-k_{\rm np}^2}$, where $k_{\rm np}$ is the nonlinear  plasma wavenumber introduced in Appendix~\ref{scatterings} and discussed in figure~12 of~\citet{luo2024control}: $k_{\rm np}\approx k_{\rm pe} -3 k_{\rm pe} (E_L/E_0)^2/16$ . If  $|k_{Ly}|<\delta k_{Ly}$, we expect  Raman scattering and wavenumber nonlinearity to dominate, with a corresponding ``quasi-longitudinal'' electric energy content $\mathcal{E}_{\rm in}$. If $|k_{Ly}|>\delta k_{Ly}$, we expect additional processes, such as the Weibel-like instability to dominate,  with electric energy content $\mathcal{E}_{\rm out}$. More specifically 
\begin{align}
     \mathcal{E}_{\rm in}=&{\int_{0.5k_{\rm pe}}^{1.5k_{\rm pe}}dk_{Lx}\int_{|k_{Ly}|<\delta k_{Ly}}}{dk_{Ly} |E_L(k_{Lx},k_{Ly})|^2}, \nonumber \\
     \mathcal{E}_{\rm out}=&{\int_{0.5k_{\rm pe}}^{1.5k_{\rm pe}}dk_{Lx}\int_{|k_{Ly}|>\delta k_{Ly}}}{dk_{Ly} |E_L(k_{Lx},k_{Ly})|^2}.
     \label{LongitudinalMode}
\end{align}

Optimally, the energy is mostly contained in the quasi-longitudinal field, while non-ideal 2D effects may scatter energy outside this region, and the ratio $\mathcal{E}_{\rm out}/\mathcal{E}_{\rm in}$ thus quantifies the importance of these 2D effects that reduce the transverse coherence of the plasma wave. In figure~\ref{Bz_energy}, the electric energies $\mathcal{E}_{in}$ and $\mathcal{E}_{out}$ are also shown, by red dashed and blue dotted lines, respectively. Most of the energy is contained in the quasi-longitudinal field, that saturates around $t \omega_{\rm pe}\approx 190$. The energy scattered outside this wavenumber region is approximately following the same trends as the $\mathcal{E}_{B_z}$ and $|H_z|_{\rm max}$ curves, with a first saturation around $t \omega_{\rm pe}\approx 190$. A more pronounced growth phase occurs during $ t\omega_{\rm pe}\approx 300$--$400$, when the magnetic fields grow due to the Weibel-like instability, followed by a second saturation. It is noteworthy that around the time of the onset of the Weibel-like instability, $\mathcal{E}_{in}$ also starts to decrease, indicating a reduced coherence of the plasma wave, {even if the ratio   $\mathcal{E}_{in}/\mathcal{E}_{out}$ is always larger than one.} 

\begin{figure}
    \centering
	\includegraphics[width=0.75\linewidth]{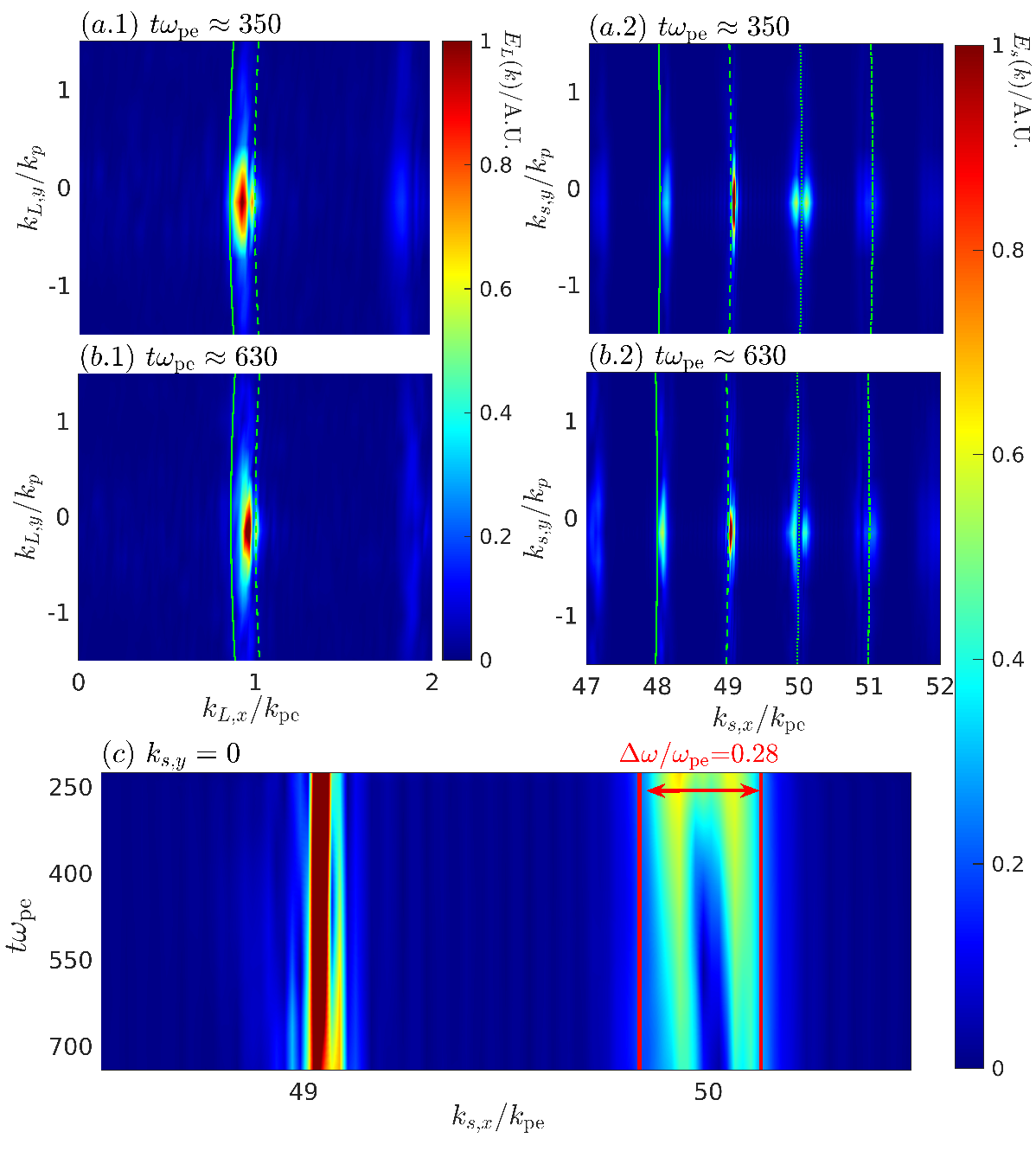}
    \caption{(a-b): The wavenumber spectra of the longitudinal electric field of the plasma wave $E_L$ (left column) and the electromagnetic wave $E_s$ (right column) at different times. The green curves correspond to the solutions of equation~(\ref{wake_fft}) (left column) and equation~(\ref{laser_fft}) (right column). In the left column, the solid (dashed) curves include (do not include) corrections for the nonlinear wavenumber shift. In the right column, different green curves represent different scattering order. (c) The temporal evolution of the purely longitudinal ($k_{s,y}=0$) wavenumber component of the main modes of the two laser beams. Vertical red lines indicate the maximum bandwidth of the first laser beam $\Delta\omega/\omega_{\rm pe}\approx 0.28$.}  
\label{fft_ExEz_later} 
\end{figure}

\subsection{Strong nonlinearity and plasma wave filamentation \label{FinalStage}}

We have established the influence of Weibel-like instability on the transverse dynamics of the plasma waves: it gives rise to transverse magnetic perturbations and a scattering of electric field energy beyond that caused by Raman scattering, potentially leading to a reduction in the coherence of plasma waves. However, the direct impact of Weibel-like instability on the laser evolution in the plasma remains negligible, {as the instability quickly saturates}. In this section, we explore how the later-stage evolution of the laser propagation in the \emph{strong-nonlinear} phase, where density modulations and filamentation arise.

Figure~\ref{fft_ExEz_later} shows the wavenumber spectra of the electric field of the plasma wave $E_L$ (left column) and that of the electromagnetic wave $E_s$ (right column) at different time instances, with the corresponding wavenumbers denoted by $k_L$ and $k_s$, respectively. Here we focus on times $t \omega_{\rm pe} \ge 350$. At early times ($t \omega_{\rm pe}=350$) the main mode spectrum of the plasma wave assumes an arc-shape between the circle given by mode matching (dashed line) and that with the nonlinear shift accounted for (solid line). By the time $t \omega_{\rm pe} = 630$ the arc-shaped spectrum is not clearly seen anymore, and a stronger transverse broadening is observed. In addition, in the vicinity of $k_{s,x}\approx 50$ of the electromagnetic wave, the slightly downshifted and upshifted components start to evolve differently, the downshifted one spreading more in the transverse direction. The Stokes/anti-Stokes scattering is barely affected.

\begin{figure}
    \centering
	\includegraphics[width=0.75\linewidth]{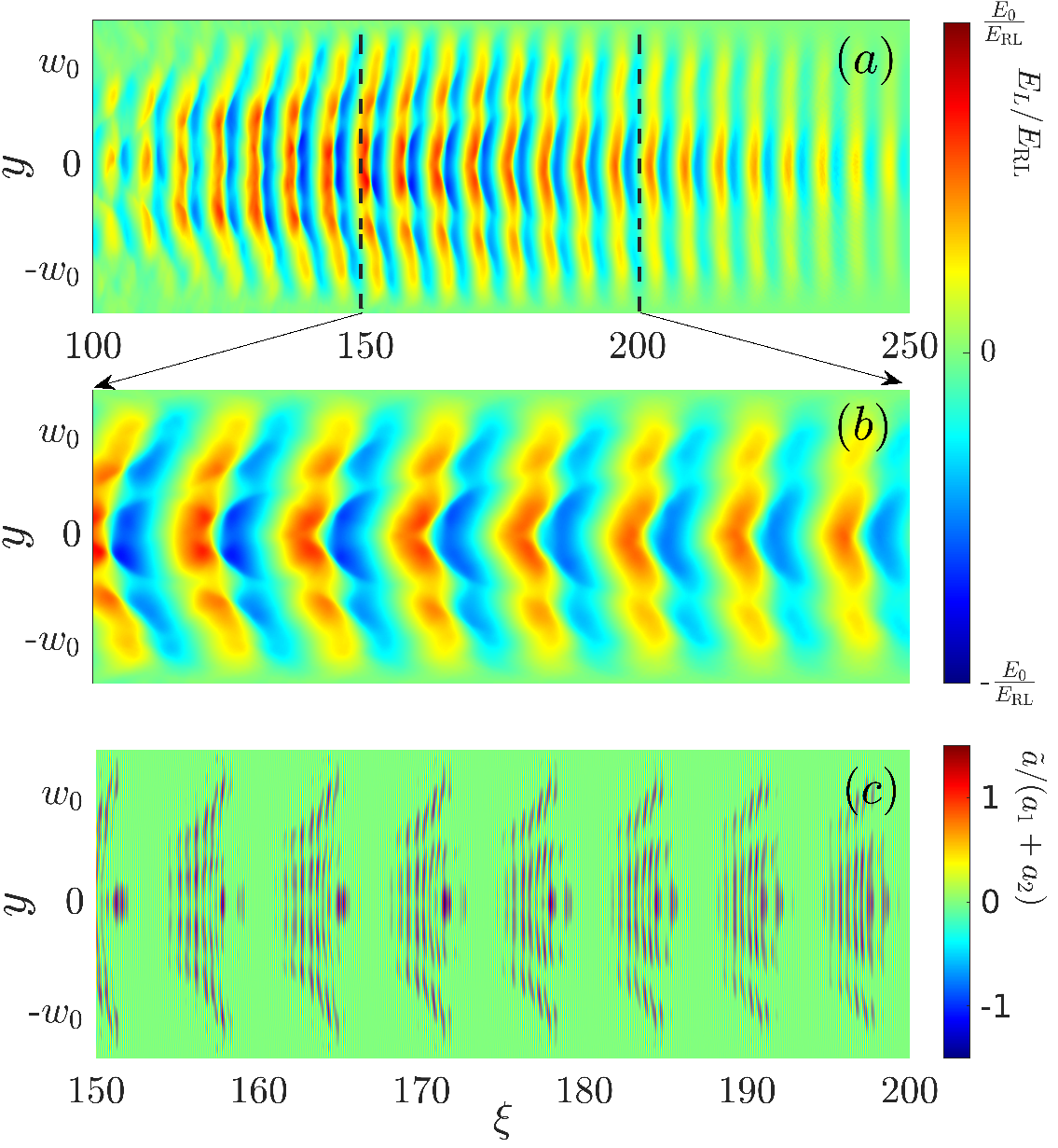}
    \caption{At $t\omega_{\rm pe}\approx 550$, (a) the electric field of the plasma wave $E_L$, over the range  $\xi\in[100,250]$, (b) $E_L$ in the range $\xi\in[150,200]$, and (c) the electric field of the electromagnetic wave $\tilde{a}$, normalized by the initial total laser amplitude $a_1+a_2$ in the range $\xi\in[150,200]$.}  
\label{wave_break} 
\end{figure}

In figure~\ref{fft_ExEz_later}(c), the temporal evolution of the wavenumber spectrum $k_{s,x}$ of the two dominant laser beams is shown for the purely longitudinal mode ($k_{s,y}=0$). The two vertical red lines indicate the maximum bandwidth of the first, chirped, laser beam -- located at $k_{s,x}/k_{\rm pe}=50$ -- that can be estimated as $\Delta\omega/\omega_{\rm pe}\approx |\alpha|T_{\rm pulse}=0.28$. Without the interaction of the lasers and the autoresonant creation of the plasma wave, the wavenumber spectrum of this chirped pulse would be nearly constant within this region. Instead, one observes that the signal in the vicinity of $k_{s,x}\approx50$ becomes very weak, leaving two bands on the sides, partly related to laser depletion due to the plasma wave excitation, also present in 1D and observed in our previous simulations,  and partly related to scattering from $k_{s,y}=0$ (plotted) to $k_{s,y}\ne 0$, as evident in figure~\ref{fft_ExEz_later}(b.2). 

\begin{figure}
    \centering
	\includegraphics[width=0.75\linewidth]{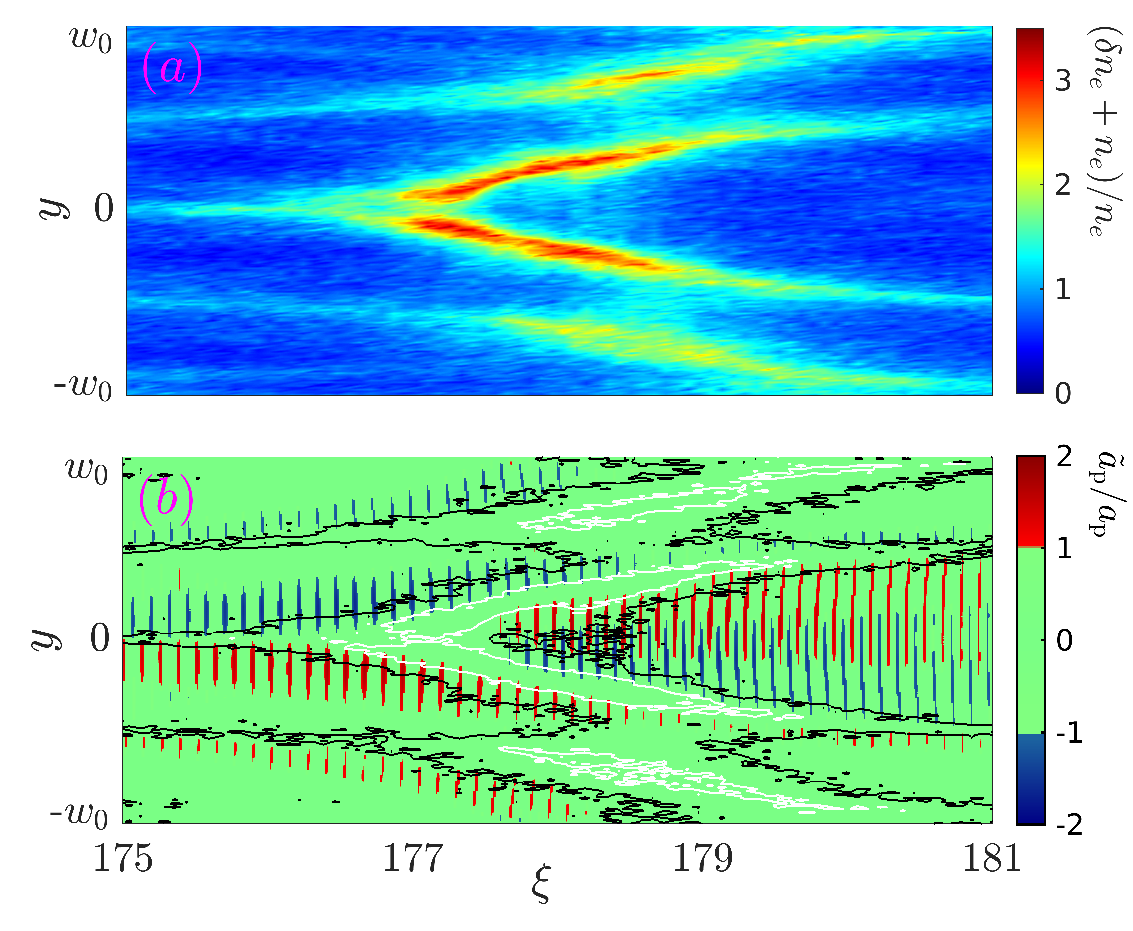}
    \caption{A low power ($a_{\rm p}=5\%a_{1,2}$) probe laser with orthogonal polarization to the lasers that create the plasma wave co-propagates together with the two dominant laser beams.
    At the time $t\omega_{\rm pe}=550$ restricted to one plasma oscillation in $\xi$, (a) the plasma wave density profile, and in (b) blue (red) corresponds to large negative (positive) values of the probe laser field, and for the purpose of visibility, values of the probe field less in magnitude than the initial peak probe field are set to zero. The black and white contours correspond to refractive index values of $\eta_r=0.9998$ and $\eta_r=0.9996$. See text for details.} 
\label{probe_density} 
\end{figure}

With the onset of autoresonance, considerable pump depletion, on the account of the plasma wave excitation, is expected to take place. In terms of frequency, the width of the depleted gap can be estimated by $\Delta\omega_{\rm eff}/\omega_{\rm pe}\approx3[1-(E_{\rm RL}/E_0)^2]/16\approx0.1$. This is around a third of the overall bandwidth of the chirped beam, consistently with the observation. This also allows us to estimate the length over which autoresonance occurs, $L_{\rm AR}\approx\Delta\omega_{\rm eff}/\omega_{\rm pe}/|\alpha|=75$. This distance $L_{\rm AR}$ is indeed representative of the spatial extent over which effective autoresonance takes place; see the  $\xi\approx 125$--$200$ range of figure~\ref{2D_electric_field}(c).

Figure~\ref{wave_break}(a-c) capture snapshots of the plasma wave and the electromagnetic field, respectively, at time $t \omega_{\rm pe}\approx 550$, the time moment between those in figure~\ref{fft_ExEz_later}(a) and figure~\ref{fft_ExEz_later}(b). In figure~\ref{wave_break}(a), the electric field carried by the plasma wave $E_L$ over the range of $\xi\in[100,250]$ is shown, where the transverse modulation is observed, particularly occurring over the range of $\xi\in[150,200]$ as exhibited in figure~\ref{wave_break}(b). The transverse regions where the beating patterns of the laser are weak, such as the gap around $|y|\approx 4$ in figure~\ref{wave_break}(c)\footnote{Note: the laser width shown in figure gives $w_0=2.4\pi$.}, are correlated with a reduced plasma wave amplitude; see the reduced amplitude transverse bands in figure~\ref{wave_break}(b). 

At this stage of the evolution, we conjecture that the appearance of transverse structure of the plasma wave and of the laser pulses will self-sustain and enhance the transverse modulation. 
To justify this, we performed an almost identical simulation, adding a co-propagating probe laser with a linear polarization orthogonal to that of the dominant laser pulses, having only $5\%$ of their amplitudes; we denote the normalized electric field of this probe beam by $\tilde{a}_{\rm p}$ and its initial value by $a_{\rm p}$ with $a_{\rm p}=5\%a_{1,2}$.

The density perturbation  of the plasma wave ($\delta n_e+n_e)/n_e$ is shown in figure~\ref{probe_density}(a), along with the electric field of the probe laser in figure~\ref{probe_density}(b), at the same time instance as in figure~\ref{wave_break}. To emphasize the impact of the density perturbation on laser field, in figure~\ref{probe_density}(b) we only plot the fields of the probe laser whenever $|\tilde{a}_{\rm p}|/a_{\rm p}\geq 1$ (these appear as blue and red spikes). The density perturbation modifies the index of refraction $\eta_r$, thereby affecting the transverse propagation of the electromagnetic wave.  We may approximate the index of refraction by
\begin{equation}
    \eta_r \approx 1-\frac{\omega_{\rm pe}^2}{2\omega_0^2}\frac{\delta n_e+n_e}{n_e\gamma}, 
    \label{RefractiveIndex}
\end{equation} 
with the Lorentz factor $\gamma$ being comparable in magnitude to $\gamma_{\perp}=(1+a^2)^{1/2}$. To show the spatial variation of the refractive index we draw the contours $\eta_r=0.9996$ (white) and $n_r=0.9998$ (black), calculated with the density variation shown in figure~\ref{probe_density}(a). As a result of the spatial variation of  $\eta_r$, the laser beams undergo fragmentation, with the various segments exhibiting different characteristics. Specifically, the intensity of the beamlets follow closely the density contours. Both the laser beams and the plasma have developed transverse modulation -- a feature not present earlier stages -- marking the transition into the \emph{strongly nonlinear} stage of the process. The evolution of the plasma wave in this stage have now a non-negligible impact on particle acceleration. The resulting reduced acceleration efficiency is apparent in figure~\ref{2D_energy_spectrum}(c), and a non-trivial transverse re-arrangement of the acceleration process is reflected in Figs.~\ref{centering} and \ref{high_energy}(b).

\section{Discussion and conclusions \label{conclusion}}

We have investigated  autoresonant plasma beat-wave excitation in 2D with the PIC simulation  {\sc Smilei}. Primary laser and plasma  parameters for the   2D study correspond to those used in a 1D simulation where large wave amplitudes and significant self-injection of electrons was observed.  A number of 2D features arise that were not previously seen using either simplified fluid models or 1D kinetic simulations. For the case studied here, these effects occur after the peak electric field is reached and at wave amplitudes beyond the RL limit. 
That is, autoresonance works in 2D in that the peak field exceeded the RL limit, but at the same time 2D phenomena limit this field to a few oscillations beyond where it obtains its peak value.  

As in 1D, electrons are self-injected into the developing plasma wave structure and are accelerated, most efficiently along the axis of the laser beams. Various scattering processes -- Raman side scattering or near-forward Raman scattering along with Stokes and anti-Stokes scattering -- excite harmonics and circular patterns in wavenumber space, although these do not have much effect on the acceleration dynamics of self-injected electrons. This is related to the relatively low value of the density chosen in this paper, $n_e/n_{cr}=0.0004$. As discussed explicitly in ~\citet{luo2024control}, Stokes and anti-Stokes scattering, present both in 1D and 2D, do not significantly affect the laser beating at this density. Similarly, the near-forward Raman scattering observed in our 2D PIC simulations does not have a significant impact on the plasma wave.

The phase fronts of the plasma wave develop curvature -- a wavefront bowing~\citep{PEM,Lin1,Lin2} -- due to the transverse variation of laser intensity and corresponding variation in the plasma wave excitation. This is seen to change the transverse location of the peak field and the energy spectrum of electrons.  Magnetic fields are self-generated by the large-scale currents driven by the density and velocity perturbation in the plasma wave~\citep{gorbunov1,gorbunov2,sheng7}. A Weibel-like instability~\citep{weibelInstability} arising from velocity space anisotropy generates transverse magnetic field perturbations. The electric field, which is initially quasi-longitudinal,  scatters in the transverse direction and the acceleration efficiency of the most energetic electrons starts to deviate from that seen in 1D. Finally, we observe transverse filamentation, and as a consequence, the acceleration efficiency is further reduced; in this regime, the most energetic electrons are found off-axis. 
 
Novel schemes to avoid diffractive and dephasing limits such as those based on Bessel beams~\citep{PhysRevAccelBeams.26.061301, np_Malka} place requirements on chirped pulses; their compatibility with autoresonant excitation is a topic for future study. Transverse effects originating from the defocusing of the laser beams may reduce acceleration efficiency, which motivates further studies utilizing a plasma channel to guide the laser beams.
In conclusion, this initial study of kinetic effects in autoresonant plasma beat-wave excitation indicates that 2D effects can become important and thus the design of any accelerator scheme based on this will require 2D analysis and optimization. The computational expense of 2D studies, supports the development of reduced models of the 2D problem, even if some of the phenomena described in this paper, such as the  Weibel-like instability -- related to a complex evolution of the electron phase space -- might be challenging to model and require further kinetic studies.

\section*{Funding}
This project received funding from the Knut and Alice Wallenberg Foundation (Grant No.~KAW 2020.0111). The computations were enabled by resources provided by the National Academic Infrastructure for Supercomputing in Sweden (NAISS), partially funded by the Swedish Research Council through grant agreement No.~2022-06725 and 2021-03943. C.~Riconda and J.~S.~Wurtele thank the Berkeley-France Fund for support of this research.

\appendix

\section{Precise control of the 1D autoresonant PBWA \label{control_duration}}

In our previous paper~\citep{luo2024control}, we have identified laser and plasma parameters that lead to an effective acceleration of self-injected electrons. These parameters are summarized in Table~\ref{table}, and the acceleration is not sensitive to small variations around these values. A sensitive parameter is the laser duration for which we found a constraint, as indicated in the last column, $T_{\rm pulse}\lesssim 100\pi/\omega_{\rm pe}$. To clearly identify the regime of autoresonant PBWA of interest for us, we conducted 1D kinetic simulations where we varied the laser duration as a key control parameter.
The other parameters were chosen as follows: normalized laser electric field $a_1=a_2 = 0.2$, a homogeneous plasma density of  $n_e/n_{cr} = 0.0004$, following a short linear density ramp, a chirp of rate $\alpha=-0.0014$ is applied to the first laser beam ($a_1$), and the frequency difference between the two lasers $\Delta\omega$ meets the resonant frequency $\omega_{\rm pe}$ at $t_0=22.5\pi$.

\begin{table}
  \begin{center}
\def~{\hphantom{0}}
  \begin{tabular}{ p{1.5cm} | p{1.5cm} | p{1.5cm} | p{1.5cm} | p{1.5cm} }
   $a_1=a_2$ & $n_e/n_{cr}$   & $\alpha$       & $t_0$     & $T_{\rm pulse}\omega_{\rm pe}$ \\[3pt]  
   $\sim0.2$ & $\sim$ 0.0004  & $\sim$ -0.0014 & 22.5$\pi$ & $\lesssim 100\pi$ \\
  \end{tabular}  
  \caption{The optimal laser and plasma parameters in the homogeneous plasma to drive autoresonant PBWA}
  \label{table}  
  \end{center}
\end{table}

\begin{figure}
    \centering
	\includegraphics[width=0.75\linewidth]{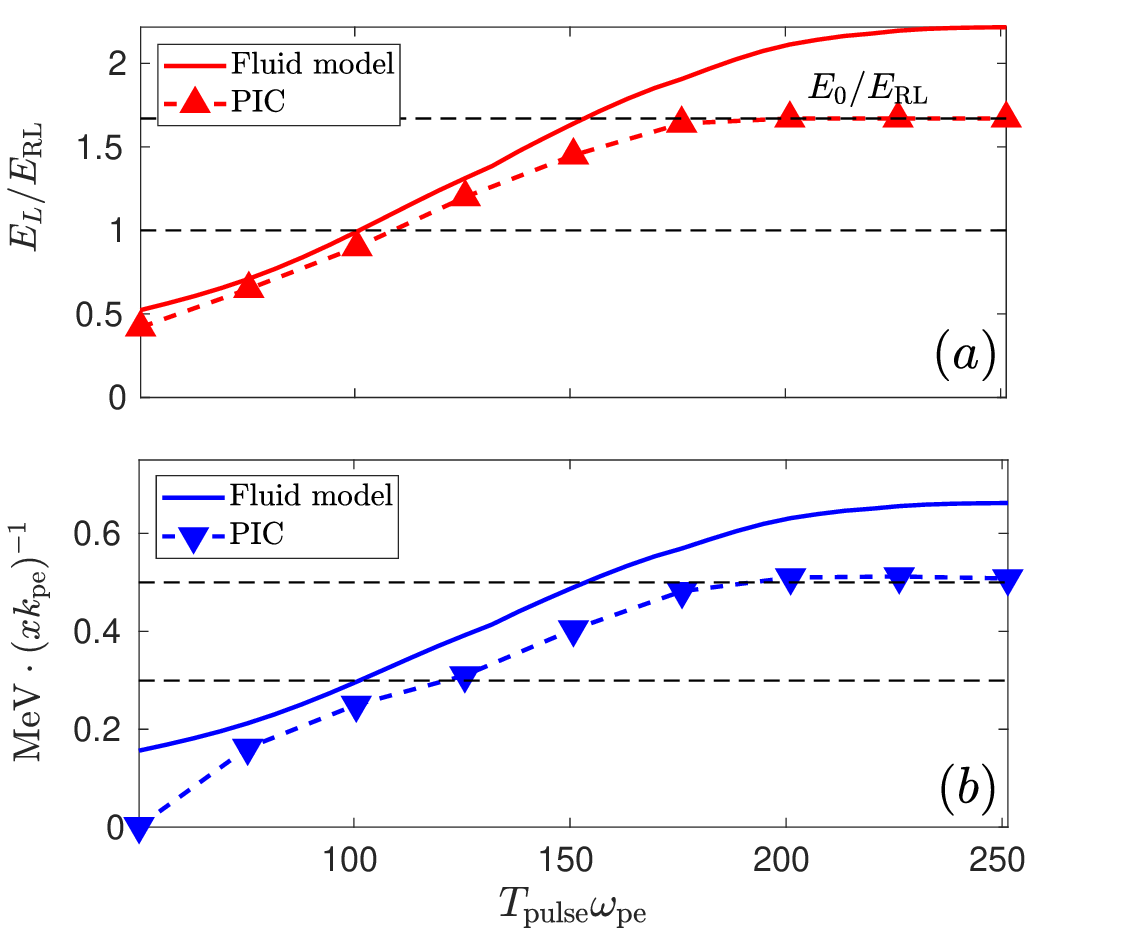}
    \caption{Comparisons  of 1D PIC and 1D fluid models for the autoresonant excitation with parameters in Table~\ref{table}. (a)  $E_L/E_{\rm RL}$. The red up-triangles represent the kinetic results, while the red line plots the results of the fluid model, as the function of laser duration $T_{\rm pulse}\omega_{\rm pe}$; (b) the maximum energy gradient in units of the normalized distance $xk_{\rm pe}$, ${\rm MeV}/(xk_{\rm pe})$, over the acceleration length $Lk_{\rm pe}\approx500$. The blue down triangles represent the kinetic results, and the blue line is an estimation based on the fluid model.}
\label{control} 
\end{figure}

In figure~\ref{control}(a), the amplitude of the plasma wave normalized by the RL limit, $E_L/E_{\rm RL}$ is shown as a function of laser duration $T_{\rm pulse}\omega_{\rm pe}$, comparing 1D kinetic simulation results (red up-triangles) and a fluid model (red curve). The two horizontal dashed black lines represent the RL limit $E_{\rm RL}$ and the wave-breaking limit $E_0$, respectively. An increase in the laser duration leads to higher amplitude plasma waves. Below the wave-breaking limit, the kinetic simulation results follow the predictions of the fluid model with reasonable agreement. However, while the electric field in the kinetic simulations represents a real upper bound for the kinetic results, which is where they saturate for $T_{\rm pulse}\omega_{\rm pe}\gtrsim 200$, the fluid predictions keep increasing with $T_{\rm pulse}$. 

In figure~\ref{control}(b) we show the energy gradient (in units of ${\rm MeV}/(xk_{\rm pe})$) experienced by the most energetic accelerated particles over the acceleration length of $Lk_{\rm pe}\approx500$, again, with a comparison of kinetic simulation results (blue down-triangles) and a fluid estimate (blue curve). The latter is calculated as $E_L m_e c^2/E_0$, where $E_L$ can be obtained from the fluid model. The kinetic simulation results cannot exceed the wave-breaking estimate, i.e., $\lesssim 0.5$. Hence, by varying the laser duration for fixed optimal laser parameters, the plasma wave amplitude and the energy gradient experienced by the accelerated particles can be well controlled. Thus, a control of the electron energization is possible, at least until the onset of multiple-dimensional effects. 

\section{Processes in the linear and weak nonlinear phases}
\subsection{Side-scattering or near-forward Raman scattering}
\label{scatterings}
 
\begin{figure}
    \centering
	\includegraphics[width=0.75\linewidth]{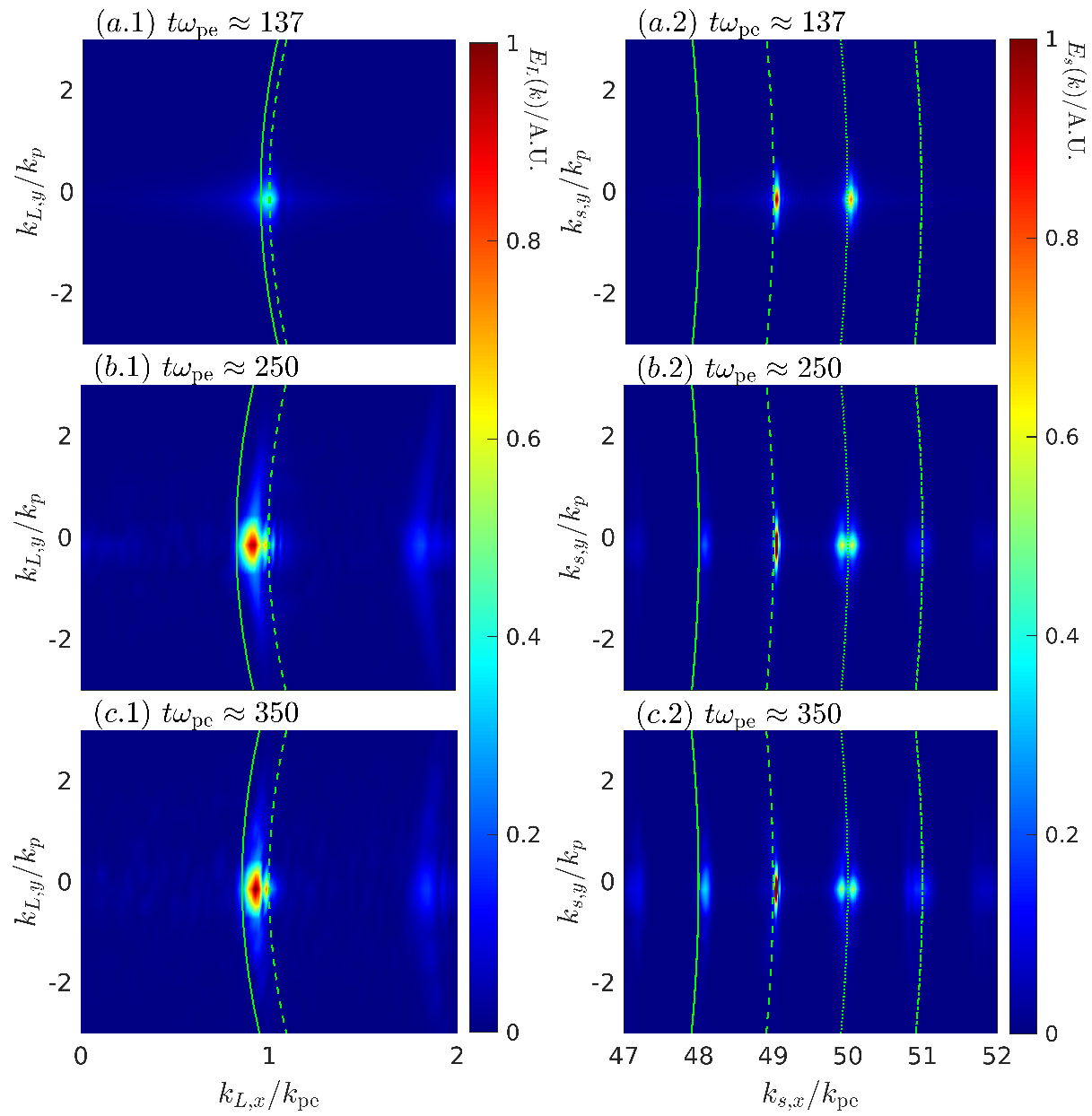}
    \caption{The Fourier spectra of the wavenumber carried by the longitudinal electric field $E_L$ (left column) and the electromagnetic wave $E_s$ (right column) are depicted at various moments, satisfying the temporal constraint $t\omega_{\rm pe}\leq350$. The green lines correspond to the solutions of equation~(\ref{wake_fft}) (left column) and equation~(\ref{laser_fft}) (right column). In the left column, the solid line takes into account the nonlinear wavelength shift due to the relativistic effect, while the dotted line neglects this effect.}  
\label{fft_ExEz_early} 
\end{figure}

 In the left column of figure~\ref{fft_ExEz_early}, we observe that a distinct signal of the plasma wave emerges during the initial linear stage, characterized by wavenumbers $k_{L,x}\approx k_{\rm pe}$ and $k_{L,y}\approx 0$, as depicted in figure~\ref{fft_ExEz_early}(a.1). Later, when autoresonance becomes significant, figure~\ref{fft_ExEz_early}(b.1), the amplitude of the plasma wave intensifies, accompanied by a reduction in wavenumber, approximately $k_{L,x}<k_{\rm pe}$. Additionally, non-zero components of $k_{L,y}$ become apparent, along with a weak harmonic signal, roughly $k_{L,x}\approx 2k_{\rm pe}$. The arc-shaped signal of the dominant wavenumber is a result of side-scattering or near-forward Raman scattering of the laser beams.

The plasma wavenumber matching condition during Raman scattering is given by
\begin{equation}
    (k_{L,x}-\delta k_L-k_0)^2+k_{L,y}^2=(k_0-\omega_{\rm pe}/c)^2. \label{wake_fft} 
\end{equation}
Here, $\delta k_L$ represents the nonlinear wavenumber shift. 
Drawing on insights from ~\citet{1990a,1990b,RevModPhys.81.1229}, we can approximate the nonlinear plasma wavelength as $\lambda_{nL}=\lambda_p(1+3(E_L/E_0)^2/16)$, where $\lambda_p$ is the linear plasma wavelength. As a result, the nonlinear plasma wavenumber can be computed as $k_{Lp}=2\pi/\lambda_{Lp}\approx k_{\rm pe}+\delta k_L$, with $\delta k_L/k_{\rm pe}=-3(E_L/E_0)^2/16$. The matching condition, equation~(\ref{wake_fft}), is indicated in figure~\ref{fft_ExEz_early} without nonlinear wavenumber shift (dashed green curve), and with the above estimate of the wavenumber shift (solid curve), where, to provide an upper bound, $E_L$ was set to be the highest electric field  observed in the simulation $E_{L,\rm max}$ to estimate $\delta k_L$. These two curves provide a reasonably accurate bound of the spectral features seen in the simulation. 

The wavenumber spectrum of the electromagnetic wave is presented in the right column of figure~\ref{fft_ExEz_early} for the same time instances. Initially, in figure~\ref{fft_ExEz_early}(a.2), the signal of the two dominant laser beams is evident. As time progresses, besides these two components, other harmonics resulting from Stokes or anti-Stokes scattering become visible, as shown in figure~\ref{fft_ExEz_early}(b.2) and (c.2). Analogously to equation~(\ref{wake_fft}), these arc-shaped signals of different components can be described by a matching condition of these scattering processes
\begin{equation}
    k_{s,x}^2+k_{s,y}^2=(k_0\pm n\omega_{\rm pe}/c)^2.
    \label{laser_fft}
\end{equation}
Here, $n$ denotes the scattering order, and equation~(\ref{laser_fft}) is plotted in the right column of figure~\ref{fft_ExEz_early} using green lines. The solutions of equation~(\ref{laser_fft}) with $n=-2$, $-1$, $0$, and $1$ are represented by the solid, dashed, dotted, and dashed-dotted lines, respectively, closely agreeing with the spectrum seen in the simulation. It is worth noting that the spectral feature of the first laser beam, i.e., $k_{s,x}\approx 50 k_{\rm pe}$, gradually splits into two pieces, from figure~\ref{fft_ExEz_early}(a.2) to figure~\ref{fft_ExEz_early}(c.2). Over the time range shown here, these two parts in the vicinity of $k_{s,x}\approx50$ remain nearly symmetrical. This peculiar feature, which later becomes asymmetric is further discussed in section~\ref{FinalStage}.

\begin{figure}
    \centering
	\includegraphics[width=0.75\linewidth]{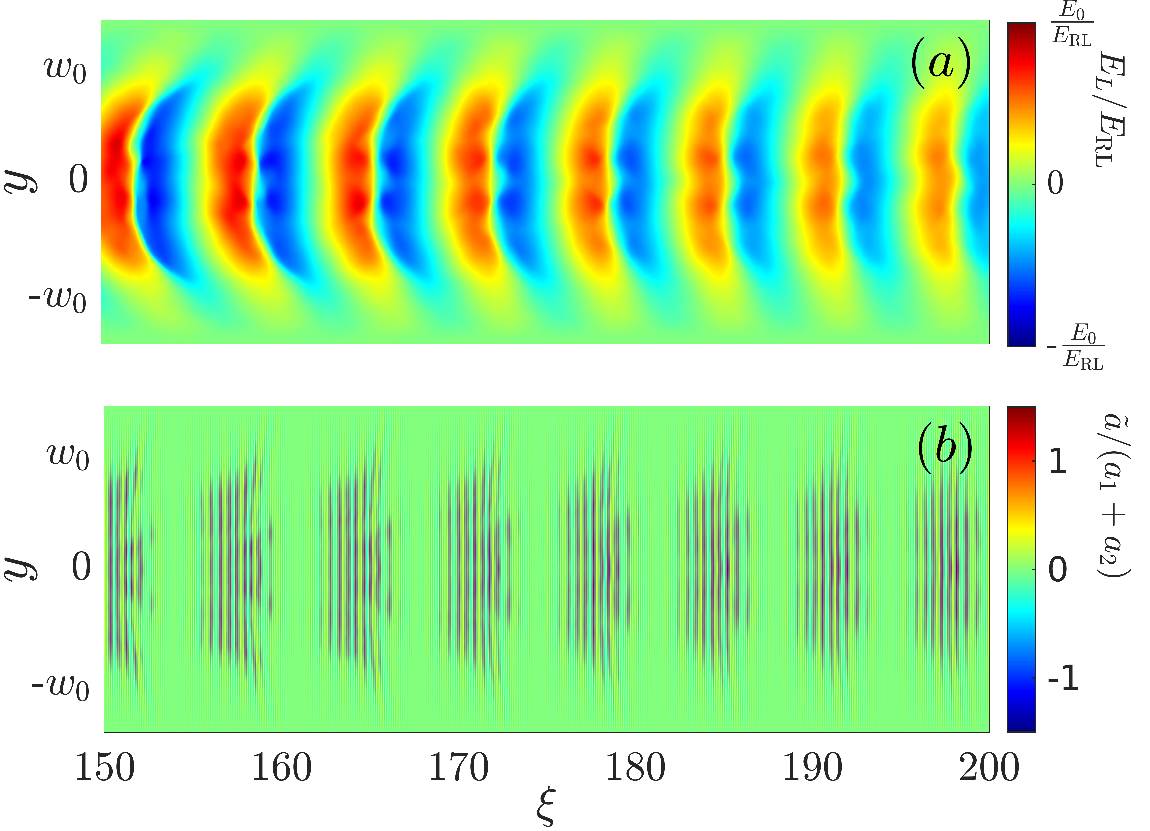}
    \caption{(a) The plasma electric field $E_L$ normalized by the RL limit  and (b) the electromagnetic wave $\tilde{a}$ normalized by the initial laser amplitude $a_1+a_2$ at the time moment $t\omega_{\rm pe}\approx350$.}  
\label{wave_bowing_early} 
\end{figure}

As discussed in ~\citet{Lin1,Lin2}, the \emph{trapped particle modulational instability} (TPMI) emerges as an alternative mechanism to cause plasma wave filamentation. This instability, whenever important, induces another wavenumber shift in addition to the one caused by the relativistic nonlinearity, and its growth rate is proportional to this shift \citep{dewar,RoseHA}. In the cases considered here, however, the wavenumber spectrum can be well described with the nonlinear shift alone (see the discussion around figure~\ref{fft_ExEz_early}), indicating that TPMI plays a negligible role.  

\subsection{Wavefront bowing of the plasma wave \label{bowing}}

Wavefront bowing emerges as a 2D effect related to the transverse variation of nonlinear wavenumber shift. In figure~\ref{wave_bowing_early}(a) and (b), the longitudinal electric field $E_L$ corresponding to the plasma wave, and the electric field $\tilde{a}_s$ of the laser beams are presented, respectively, at the time moment of $t\omega_{\rm pe}\approx 350$, corresponding to the time of figure~\ref{fft_ExEz_early}(c.1) and (c.2).  In figure~\ref{wave_bowing_early}(a), significant wavefront bowing is observed within the range of $\xi\in[150,180]$. The primary reason for this is the transverse dependence of the laser amplitude. The higher plasma wave amplitude on the middle of the laser beams corresponds to a longer wavelength, thus the wave front in this region lags behind that at the edges of the laser beam. Notably, the beating pattern of the two laser beams, shown in figure~\ref{wave_bowing_early}(b), remains regular, indicating that the influence of the density perturbation on the laser propagation remains negligible at these early times.

In conclusion, during the \emph{linear} and \emph{weak-nonlinear} phases, $t\omega_{\rm pe}\le350$, the structure of the autoresonant plasma wave remains stable and regular, as supported by figure~\ref{2D_electric_field}, despite the observed wavefront bowing in figure~\ref{wave_bowing_early}. Consequently,  the acceleration process of energetic electrons in 2D 
 remains similar to that in the 1D case, as depicted in figure~\ref{2D_energy_spectrum}(c). At this time the majority of these particles are concentrated centrally in the transverse direction, as evident in figure~\ref{centering} and figure~\ref{high_energy}(a).

\bibliographystyle{jpp}

\bibliography{reference}

\end{document}